# On PSR1913+16

A. LOINGER

Dipartimento di Fisica, Università di Milano

Via Celoria, 16 – 20133 Milano, Italy

**Summary.** – There are reference frames for which *both* stars of binary radio-pulsar PSR1913+16 are at rest. As a consequence, PSR1913+16 does not emit gravity waves. (If the accelerated bodies should send out a gravitational radiation, they would represent a class of physically privileged systems. But this is forbidden by general relativity.)

**1**. – I have already discussed critically in previous papers [1], though *en passant*, the question of the gravitational waves which – as it is commonly believed – would be sent forth by the binary pulsar PSR1913+16. I wish now to point out a drastic argument against the possibility that the observational data can be interpreted as an indirect proof of a real emission of gravity waves from PSR1913+16.

**2**. – I begin with some consideration on the role of the reference frames in general relativity.

At p.268 of Weyl's book *Raum-Zeit-Materie* [2] we read: "Zunächst stellen wir fest, daß *der Begriff der Relativbewegung zweier Körper gegeneinander in der allgemeinen Relativitätstheorie ebensowenig einen Sinn hat wie der Begriff der absoluten Bewegung eines einzigen*. […] Wie die beiden Körper sich auch bewegen mögen, immer kann ich durch Einführung eines geeigneten Koordinatensystems sie beide zusammen auf Ruhe transformieren."

Let us examine, however, the case of the Earth and of the sky of the "fixed" stars. Weyl remarks (*loc. cit.* in [2], p.269): "Wohl ist es unberechtig zu sagen, daß die Erde sich relativ zu den Fixsternen drehe; aber sie dreht sich in bezug auf denjenigen Körper, dem am Ort $O$ der Erde selbst gebildet wird von den Lichtstrahlen, die in $O$ von den Fixsternen her zusammenkommen. Das ist ein wesentlicher Unterschied, weil die Lichtstrahlen abhängig sind von dem metrischen Felde, das zwischen der Erde und den Fixsternen herrscht. Wir wollen auch diesen Sternenkompaß [*star compass*] genau beschreiben."

Actually, Weyl demonstrates that we can assert that the Earth rotates with respect to the above *star compass*, provided that *we suppose that the pseudo-Euclidean metric reign at infinity*. More generally, Fock [3] tries to specify an entire class of physically privileged frames – the harmonic systems, together with the requirement of pseudo-Euclidean metric at infinity – such that Galilei's motto "E pur si muove!" is literally true. As a matter of fact, however, *no class* of privileged frames can exist in general relativity, as it was repeatedly emphasized by Einstein. This means that Weyl's sentences, which we have written at the beginning of this section, have a strict sense, which does not allow any exception or attenuation.

**3**. – *Accordingly, let us choose a reference frame for which **both** stars of PSR1913+16 are at rest*. **Evidently, an observer Ω dwelling in such a system does not record any emission of gravity waves**.

**4**. – The overwhelming majority of the astrophysicists think that the temporal decrease of the revolution period of PSR1913+16 should be ascribed to an emission of gravity waves. Diligent people computed orbital data and masses by means of the semi-*empirical* PPN-method, then utilized some relativistic *perturbative* formulae. However, the reliability of the results is dubious because, owing to objective difficulties, the employed procedure is rather remote from a precise treatment of the real question, which would imply an *accurate* solution of the relativistic problem of two bodies of finite masses (the two stars of PSR1913+16). On the other hand, by virtue of the proposition of sect.**3**, the exact solution cannot yield any gravitational wave. Moreover, I observe that if the accelerated bodies should emit gravity waves, *the accelerated frames would be physically privileged*. But this is forbidden by general relativity.

*An objection*: An observer like $\Omega$, who dwells in a frame for which two revolving electric charges are at rest, "sees" the propagation of electromagnetic waves. ***Answer***: Of course. But the existence of the e.m. waves is a consequence of the fact that Maxwell theory (which includes the material equations $\boldsymbol{D} = \varepsilon \boldsymbol{E}$ and $\boldsymbol{H} = (1/\mu)\boldsymbol{B}$) is, *first of all*, the theory of the e.m. field ($\boldsymbol{E}, \boldsymbol{B}$) in a *Minkowski* space-time: the wave-like character of a given e.m. perturbation is a property valid for all the Galilean systems. The Riemannian formulation is only an extension.

    Historically, the belief in the physical existence of gravity waves had its origin in the linearized approximation of Einstein equations, which is a rough theory of "weak" gravitational fields in a given *Minkowski* space-time. By formal analogy, it seemed also that any accelerated mass point should generate a gravitational radiation. Then, these convictions were extended to perturbative approximations of higher orders. However, the exact non-linear theory does not admit any class of physically privileged frames for which, *in particular*, the wave-like character of a gravity field is an invariant property.

**5**. − Many physicists believe that the good accordance of the observational data concerning the PSR1913+16 with the numerical results yielded by PPN-method plus perturbative formulae is a confirmation of the theoretical validity of the procedure. But this is an epistemological mistake: an empirical agreement does not imply a conceptual adequacy. A convincing example: by means of a clever use of cycles and epicycles the followers of the Ptolemaic system calculated the planetary orbits with a *very good* accuracy (with the only exception of Mercury's orbit).

**6**. − In the last twenty years , several detailed papers on perturbative treatments of Einstein equations *in vacuo*, $R_{ik} = 0$, have been published. It seems that if the background space-time possesses symmetries, and the perturbations respect them, the perturbative approach is misleading. However, without the constraints represented by symmetries, the perturbation theory seems to be reliable.

As it is *a priori* **obvious** from the theory of the characteristics by Levi-Civita, there exist solutions which are formally of an undulatory type, but any wave-like character is illusive [1].

PARTHIAN'S ARROW

Let us suppose that our observer $\Omega$ (see sect.**3**) is a believer in the existence of the gravity waves. He wishes to calculate the watts of gravitational radiation emitted by the rest of the universe, which revolves around him. $\Omega$ is a learned man, he reads at pp.359 and 360 of Fock's treatise [3] that, as a consequence of a well-known approximate formula the order of magnitude of the power of the gravity radiation sent out by Jupiter, e.g., in its motion around the Sun is given by $B^2/(2 \cdot 10^{39} c^4)$ g·s$^{-1}$, where $B \equiv m\omega v^2$; $m \cong 2 \cdot 10^{30}$ g is Jupiter's mass;

$\omega \cong 2 \cdot 10^{-8}$ s$^{-1}$ is Jupiter's angular velocity of revolution; $v^2/c^2 \cong 2 \cdot 10^{-9}$, if $v$ is Jupiter's mean linear velocity, ($v = \omega r$). Substituting these numerical values, we obtain $B^2/(2 \cdot 10^{39} c^4) \cong 5 \cdot 10^{-12}$ g·s$^{-1} \triangleq 450$ watt. Fock's remarks that the power of the solar electromagnetic radiation is $4 \cdot 10^{12}$ g·s$^{-1}$, i.e. about $10^{24}$ times greater.

The observer $\Omega$ imitates Fock's computation and substitutes in Fock's formula some reasonable estimates of the total mass of the world and of its "mean" radius; for $\omega$, of course, no problem. $\Omega$ is a very rich man and possesses Michelson interferometers with kilometric arms and suspended mirrors. Do you think that he will be successful in detecting the gravitational radiation which – in accordance with his belief – should be emitted by the rest of the universe?

(Our hero $\Omega$ had read in Pauli's book [4] that Einstein equations with cosmological term admit of a unique solution *in vacuo*, which is regular everywhere: $g_{ik} \equiv 0$. Now – as Pauli remarked – this implies Mach's relativity of the inertial forces, in particular that a "purely kinematical" rotation does *not* exist.)